\documentclass[9pt,twocolumn,twoside]{osajnl}

\journal{arxiv} 

\setboolean{shortarticle}{false}

\usepackage{lineno}

\usepackage{amsmath}
\usepackage{graphicx}

\usepackage[absolute]{textpos}
\usepackage{xcolor}

\begin{document}

\title{Characterizing non-polarization-maintaining highly nonlinear fiber toward squeezed-light generation}

\author[1,*]{Joseph C. Chapman}
\author[1]{Nicholas A. Peters}

\affil[1]{Quantum Information Science Section, Oak Ridge National Laboratory, Oak Ridge, TN 37831}
\affil[*]{chapmanjc@ornl.gov}


\begin{abstract}
Squeezed light, which is easily degraded by loss, could benefit from generation directly in optical fiber. Furthermore, highly nonlinear fiber could offer more efficient generation with lower pump power and shorter fiber lengths than standard single-mode fiber. We investigate non-polarization-maintaining highly nonlinear fiber (HNLF) for squeezed-light generation by characterizing possible sources of excess noise, including its zero-dispersion wavelength (ZDW) variation and polarization noise. We find significant ZDW variation and excess polarization noise.  We believe the polarization noise is from non-linear polarization-mode dispersion. We model this polarization noise and find that it is likely to degrade Kerr squeezing but not squeezing from four-wave mixing.
\end{abstract}

\setboolean{displaycopyright}{false}


\maketitle

\begin{textblock}{13.3}(1.4,15)\noindent\fontsize{7}{7}\selectfont\textcolor{black!30}{This manuscript has been co-authored by UT-Battelle, LLC, under contract DE-AC05-00OR22725 with the US Department of Energy (DOE). The US government retains and the publisher, by accepting the article for publication, acknowledges that the US government retains a nonexclusive, paid-up, irrevocable, worldwide license to publish or reproduce the published form of this manuscript, or allow others to do so, for US government purposes. DOE will provide public access to these results of federally sponsored research in accordance with the DOE Public Access Plan (http://energy.gov/downloads/doe-public-access-plan).}\end{textblock}
\section{Introduction}
Squeezed light is a valuable resource to increase signal-to-noise ratio beyond the standard quantum limit and enable continuous-variable (CV) quantum entanglement~\cite{bachor2019guide}.  Squeezed light has been used to enhance the sensitivity of many different types of sensors~\cite{QSRev}. Most notably, it has been demonstrated to enhance the sensitivity of interferometers, aiding in the detection of gravitational waves~\cite{Aasi2013}. It is also used as a fundamental resource in CV quantum-computing and quantum-repeater proposals~\cite{RevModPhys.77.513,RevModPhys.84.621}. The first demonstration of squeezed light was measured in sodium vapor in 1985~\cite{PhysRevLett.55.2409}. Shortly following, several demonstrations using optical fiber showed similar amounts of squeezing~\cite{PhysRevLett.57.691,PhysRevLett.59.2566}. Due to the relatively low non-linearity of standard optical fiber, early demonstrations required hundreds of mW of pump power to produce significant squeezing~\cite{Bergman:91,Bergman:94}. A fiber with higher nonlinearity could reduce the necessary pump power for more efficient squeezed-light generation and allow for shorter fibers to reduce the dependence on fiber variations.

Highly nonlinear fiber~\cite{Sudo1990,secondHNLF,ONISHI1998204,806765,1561390,1561391} is a dispersion-shifted doped-core silica fiber with a large refractive-index difference between the core and cladding. The doping and large refractive index difference enable a larger non-linearity~\cite{HNLFreview}. There are four different types of HNLF~\cite{HNLFreview}. Type-I HNLF has a small and anomalous dispersion in the operation window, thus optical solitons are easily generated. Type-II and -III HNLF are dispersion-shifted so the zero-dispersion wavelength lies in the operation window. This allows for efficient four-wave mixing~\cite{1016354,el_20030544} and wavelength conversion~\cite{632764}. Type-III HNLF has almost zero dispersion slope and, as a result, is called dispersion-flattened HNLF, which comes at the cost of a lower nonlinearity compared to Type-II.  Type-III HNLFs have been used for supercontinuum generation~\cite{el_20030599} and ultra-short pulse compression~\cite{Tamura:01}. Type-IV have large normal dispersion and find use in Raman amplification~\cite{1561376} and pulse compression via linear chirping~\cite{1650529}. Type-II HNLF is what we have~\cite{doi:10.1063/1.5048198} and used in our tests.

Squeezed light can be degraded by the presence of excess noise. Zero-dispersion wavelength variation, Raman scattering, and non-linear polarization-mode dispersion are possible sources of excess noise that can be produced by non-polarization-maintaining HNLF. In this work, after a review of relevant background and some general analysis of squeezing with added noise, we characterize several lengths of non-polarization-maintaining HNLF by measuring the zero-dispersion wavelength and its variation. We also measure the polarization noise produced by several lengths of non-polarization-maintaining HNLF and compare it to regular single-mode and polarization-maintaining fiber. Since polarization noise is found to be non-negligible, we also develop a model for how it can affect several types of squeezing.
\section{Relevant squeezing overview}
\subsection{Gaussian state notation}
To begin our discussion of squeezing methods, we first briefly introduce our notation we will use throughout. Following the convention of Ref.~\cite{RevModPhys.84.621}, Gaussian states are fully characterized by their displacement vector $\Bar{\textbf{x}}$ and their covariance matrix $\textbf{V}$. 
\begin{equation}
    \Bar{\textbf{x}}:=\text{Tr} (\Hat{\textbf{x}}\rho),
\end{equation}
where $\Hat{\textbf{x}}=(\Hat{q}_1,\Hat{p}_1,...,\Hat{q}_N,\Hat{p}_N)^T$ and the set $\{\Hat{q}_k,\Hat{p}_k\}_{k=1}^N$ are the quadrature field operators for $N$ modes. The covariance matrix is defined by
\begin{equation}
    V_{ij}:=\frac{1}{2}\langle \{\Delta \Hat{x}_i, \Delta \Hat{x}_j \}\rangle,
\end{equation}
where $\Delta \Hat{x}_i:=\Hat{x}_i-\langle \Hat{x}_i \rangle$ and \{,\} is the anticommutator. In this language, a Gaussian unitary is characterized by the following transformations:
\begin{equation}
    \Bar{\textbf{x}} \rightarrow \textbf{S}\Bar{\textbf{x}}+\textbf{d}\text{, } \textbf{V}\rightarrow \textbf{S}\textbf{V}\textbf{S}^T, \label{eq:transforms}
\end{equation}
where $\textbf{d}\in \mathbb{R}^{2N}$ and $\textbf{S}$ is a $2N\times 2N$ real matrix. To preserve the bosonic commutation relations (with $\hbar=2$), $\textbf{S}$ must be symplectic, i.e.,
\begin{equation}
    \textbf{S}\mathbf{\Omega}\textbf{S}^T =\mathbf{\Omega},
\end{equation}
where 
\begin{equation}
    \mathbf{\Omega}=\overset{N}{\underset{k=1}{\oplus}}\boldsymbol{\omega}=\begin{pmatrix}
    \boldsymbol{\omega} & &\\
    & \ddots &\\
    & & \boldsymbol{\omega}\\
    \end{pmatrix}\text{, }\boldsymbol{\omega}:=\begin{pmatrix}
   0&1\\
   -1&0\\
    \end{pmatrix},
\end{equation}
known as the symplectic form. In this convention, shot noise has a variance of 1. For example, a beamsplitter between $N=2$ modes produces the symplectic map
\begin{equation}
    \textbf{B}({T}):=\begin{pmatrix}
    \sqrt{{T}}\textbf{I} & \sqrt{1-{T}}\textbf{I}\\
    -\sqrt{1-{T}}\textbf{I} &  \sqrt{{T}}\textbf{I}\\
    \end{pmatrix},\label{eq:BS}
\end{equation}
where $\textbf{I}$ is the $2\times2$ identity matrix and ${T}=1/2$ corresponds to a balanced beamsplitter. Homodyne detection measures the rotated quadrature of the signal rotated by $\theta$, the phase difference between the LO and signal. This rotation can be modeled by the simplectic map
\begin{equation}
    \textbf{R}(\theta):=\begin{pmatrix}
    cos\theta & sin\theta \\
    -sin\theta & cos\theta\\
    \end{pmatrix}.
\end{equation}

\subsection{Concise review of related fiber-based squeezing methods}
The ways to generate squeezing in optical fiber center around the two primary nonlinear interactions used, four-wave mixing and the Kerr effect, both of which depend on the 3rd-order non-linearity of the fiber~\cite{bachor2019guide}. Four-wave mixing is a multi-photon interaction that requires momentum conservation of the photons to be satisfied, also known as phasematching~\cite{LevensonNLO}. In the case of degenerate four-wave mixing where the signal is at the zero-dispersion wavelength, the quadratures undergo the symplectic map
\begin{equation}
    \textbf{S}_{\text{FWM}}=\begin{pmatrix}
    \sqrt{G}+\sqrt{G-1} & 0\\
    0 & \sqrt{G}-\sqrt{G-1}\\
    \end{pmatrix},
\end{equation}
where $G=1+(\gamma P_p L)^2$ is the phase-insensitive gain, $\gamma$ is the non-linearity coefficient, $P_p$ is the pump power, $L$ is the fiber length. This can be easily derived from the phase-dependent gain of a degenerate one-mode parametric amplifier~\cite{Andrekson:20}.

In contrast, the Kerr effect simply describes an intensity-dependent refractive index and can give rise to self-phase modulation~\cite{LevensonNLO} at any optical wavelength that propagates down the fiber. The Kerr effect produces the symplectic map
\begin{equation}
    \textbf{S}_{K}=\begin{pmatrix}
    1 & 0\\
    2 r_K & 1\\
    \end{pmatrix},
\end{equation}
where $r_K=\gamma P_p L$~\cite{bachor2019guide,Anashkina:20}.

Four-wave mixing in fibers has achieved squeezing up to about 5 dB~\cite{PhysRevLett.94.203601}. Seeded non-degenerate four-wave mixing using intensity difference detection in rubidium vapor has produced similar levels of squeezing with a relatively simple experimental setup that does not require a local oscillator~\cite{McCormick:07}. Four-wave mixing based squeezing requires gains $>1$ which can be difficult to achieve without other issues arising, as the first four-wave mixing demonstration reported~\cite{PhysRevLett.55.2409}. Moreover, the fiber dispersion constrains the wavelengths that efficiently generate four-wave mixing. For optimal phasematching, the pump should be at the zero-dispersion wavelength (ZDW) of the fiber~\cite{mechels1997accurate} but under certain conditions, phase matching can also be achieved when the pump is in the anomalous dispersion regime.

Squeezing from the Kerr effect occurs due to a large non-linear phase shift at high intensity. This is normally done with high-peak-power laser pulses to raise the stimulated Brillouin scattering threshold, which is lower for high-power continuous-wave laser light~\cite{Bergman:91}. This method was demonstrated using a symmetric Sagnac interferometer that measured squeezing up to about 5 dB~\cite{Bergman:91,Bergman:94} and is limited by frequency chirp from dispersion~\cite{Shirasaki:90}. To minimize the effects of dispersion, solitons, pulses that propagate without temporal or spectral distortion, allow this squeezing to occur and be measured more efficiently with a local oscillator~\cite{Shirasaki:90,PhysRevLett.66.153,bachor2019guide}. To produce a soliton of sufficient energy that it can be also used as a local oscillator, the central wavelength must be far enough away from the ZDW where there is enough dispersion to balance the non-linear phase shift, e.g., the closer to the ZDW, the less energy needed for soliton formation and vice versa. It has been theorized that soliton pulses enable higher amounts of squeezing compared to non-soliton pulses when using a Sagnac interferometer with a balanced beamsplitter~\cite{Shirasaki:90}. Interestingly, the proposal of Ref.~\cite{Shirasaki:90}, can be also interpreted as arising from degenerate four-wave mixing when pumped at the ZDW~\cite{Shirasaki:91,Marhic_1991}. As an alternative to the symmetric Sagnac proposal, it has been shown that an asymmetric soliton Sagnac of certain fiber lengths can produce directly detectable squeezing~\cite{PhysRevLett.81.2446,Levandovsky:99}.  However, since our pump lasers and the ZDW of our HNLF is in the 1550-nm region, we cannot generate high-power solitons for squeezing and so we are instead interested in the symmetric Sagnac demonstrations.  {In this work, we focus on non-polarization-maintaining HNLF, despite the fact that previous demonstrations of fiber squeezing have exclusively used polarization-maintaining fiber. We do this primarily because of the significantly greater cost-effectiveness of non-polarization-maintaining HNLF and we think it would be preferable in principle to not need a polarization-maintaining fiber so we wanted to investigate that.} Here, we have briefly reviewed the Kerr squeezing literature; a more detailed review can be found in Ref.~\cite{SIZMANN1999373}.

\subsection{Squeezing and excess noise}
Here we model the effects of multiple types of excess noise on several examples of squeezing. The first type of excess noise to consider is from the pump laser itself. To model excess pump laser noise using the formalism above, the input covariance matrix [\eqref{eq:transforms}] is no longer the identity (as in the case of a shot-noise limited pump laser) but the quadrature(s) with excess noise are increased accordingly.
\begin{equation}
    \textbf{V}_\textbf{p}=\begin{pmatrix}
    e_I & 0\\
    0 & e_P\\
    \end{pmatrix},\label{eq:pump}
\end{equation}
where $e_I$ ($e_P$) is the excess pump intensity (phase) noise.
To model the added noise in the channel or system, we use the model for a Gaussian channel~\cite{PhysRevA.63.032312,RevModPhys.84.621}
\begin{equation}
    \Bar{\textbf{x}} \rightarrow \textbf{T}\Bar{\textbf{x}}+\textbf{d}\text{, } \textbf{V}\rightarrow \textbf{T}\textbf{V}\textbf{T}^T+\textbf{N},
\end{equation}
 where $\textbf{T}=\textbf{I}$, $\textbf{I}$ is the identity matrix, $n_I$ ($n_P$) is the intensity (phase) noise contribution and
 \begin{equation}
     \textbf{N}=\begin{pmatrix}
     n_I & 0\\
     0 & n_P\\
     \end{pmatrix}.\label{eq:N}
 \end{equation}
 
 Let us now combine these noise sources with several types of squeezing. In the case of four-wave mixing, we have
 \begin{equation}
     \textbf{V}_{\text{FWM}}=\textbf{T}\textbf{S}_{\text{FWM}}\textbf{V}_\textbf{p}\textbf{S}_{\text{FWM}}^T\textbf{T}^T+\textbf{N},
 \end{equation}
 where in matrix form it is
 \begin{multline}
     \textbf{V}_{\text{FWM}}=\\
     \begin{pmatrix}
     e_I(\sqrt{G}+\sqrt{G-1})^2+n_I & 0\\
     0 & e_P(\sqrt{G}-\sqrt{G-1})^{{2}}+n_P\\
     \end{pmatrix}.
 \end{multline}
 In this simple example, we see how the phase squeezing produced by FWM can be reduced by phase noise of either type but not by intensity noise.
 
 For Kerr squeezing, we have
  \begin{equation}
     \textbf{V}_{\text{K}}=\textbf{T}\textbf{S}_{\text{K}}\textbf{V}_\textbf{p}\textbf{S}_{\text{K}}^T\textbf{T}^T+\textbf{N},
 \end{equation}
 where in matrix form it is
 \begin{equation}
     \textbf{V}_{\text{K}}=\begin{pmatrix}
     e_I+n_I & 2 e_I r_K\\
     2 e_I r_K & e_P+n_P+4 e_I r_K^2\\
     \end{pmatrix}.\label{eq:noise}
 \end{equation}
Kerr squeezing is not in the intensity or phase quadrature directly but the optimal squeezing is found when rotating the quadratures by $\theta_{sq}$ using a rotation matrix $\textbf{R}(\theta)$. $\theta_{sq}=-\arctan(-1/r_K)/2$ when there is no excess noise. Using that angle as an estimate, the covariance matrix transforms to $V_{sq}=\textbf{R}(\theta_{sq})\textbf{V}_{\text{K}}\textbf{R}(\theta_{sq})^T$. For illustration, the intensity quadrature variance (upper left matrix element) is 
\begin{align}
    V_{sq}[1,1]=&(e_I+n_I)\cos^2(\frac{1}{2}\arctan(\frac{1}{r_K}))-\frac{2e_I}{\sqrt{1+\frac{1}{r_K^2}}}\notag\\
    &+(e_P+n_P+4e_I r_K^2)\sin^2(\frac{1}{2}\arctan(\frac{1}{r_K})).
\end{align}

Different from FWM, Kerr squeezing is sensitive to all types of noise. The squeezing angle {$\theta_{sq}=-\arctan(-1/r_K)/2$ is always less than or equal to 45$^{\circ}$ so Kerr squeezing will be more sensitive to intensity noise since the intensity quadrature is at 0$^{\circ}$}. Rotations and beamsplitters can cause mixing between different types of noise and different types of squeezing have varying noise sensitivities, thus it is best to exactly model the particular system of interest to determine its sensitivity to different types of noise.

\section{Highly nonlinear fiber characterization for squeezing}
\subsection{Zero-dispersion wavelength}
\begin{figure}
\centerline{\includegraphics[width=1\columnwidth]{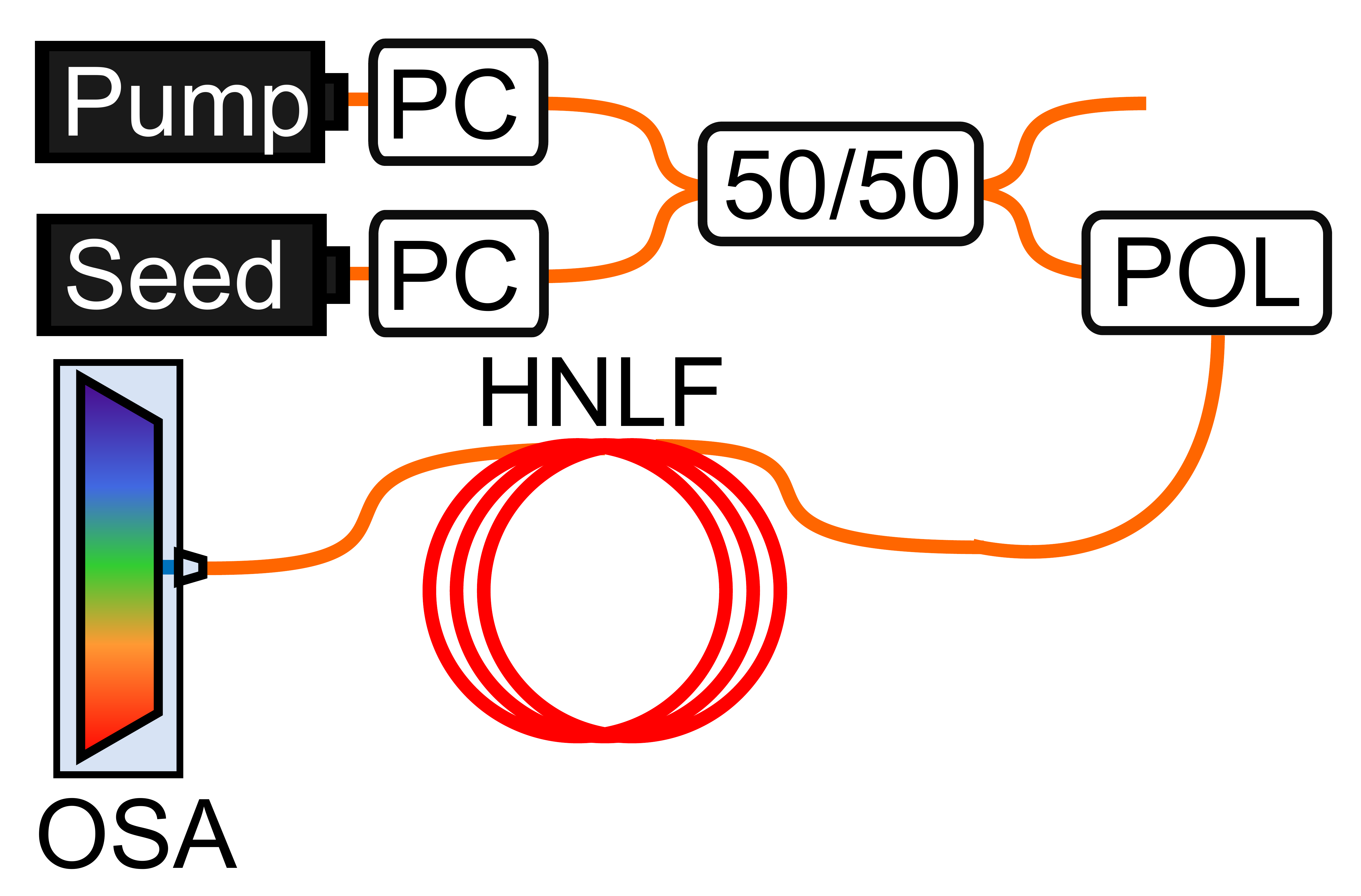}}
\caption{Zero-dispersion-wavelength measurement setup.  The pump and seed lasers (separated by 4 nm) are combined and co-polarized, then we send them through HNLF. Afterwards the pump, seed, and conjugate powers are recorded from the OSA.  HNLF = highly nonlinear fiber. OSA = optical spectrum analyzer. PC = polarization controller.}
\label{fig:HNLFZDWsetup}
\end{figure}

To characterize the average ZDW of the HNLF that we have (OFS HNLF-SPLINE ZDW), we followed a similar method to Ref.~\cite{mechels1997accurate}. The method is predicated on four-wave mixing being most efficient when the pump is at the ZDW because there is the optimal phasematching. To find the ZDW of the fiber, we use the setup shown in Fig.~\ref{fig:HNLFZDWsetup}.  Two tunable continuous-wave (CW) external-cavity lasers (Pure Photonics PPCL550) are combined and co-polarized in single-mode fiber. We set the wavelength spacing to be 4-nm apart and left the spacing constant as we discretely swept the pump (the longer of the two wavelengths) and the seed through the estimated ZDW of 1542 nm. We did this for our 98-m and 510-m spool of HNLF. At each wavelength setting, we recorded the power of the pump, seed, and conjugate (power at 4-nm higher than the pump) from the optical spectrum analyzer (OSA, Yokogawa AQ6370B). The pump and seed powers, are each around 0 dBm, with the pump being about 1-dB larger, to have enough power to measure the conjugate power on the OSA but not too powerful so the pump and seed saturate the OSA. To divide out power fluctuations of the pump, we normalized the conjugate power ($P_c$) by the seed power ($P_s$) and square of the pump power ($P_p$) since with approximation $P_c=P_s G$, where the gain ($G$) is proportional to $P_p^2$~\cite{Wang_2001}.

\begin{figure}
\centerline{\includegraphics[width=1\columnwidth]{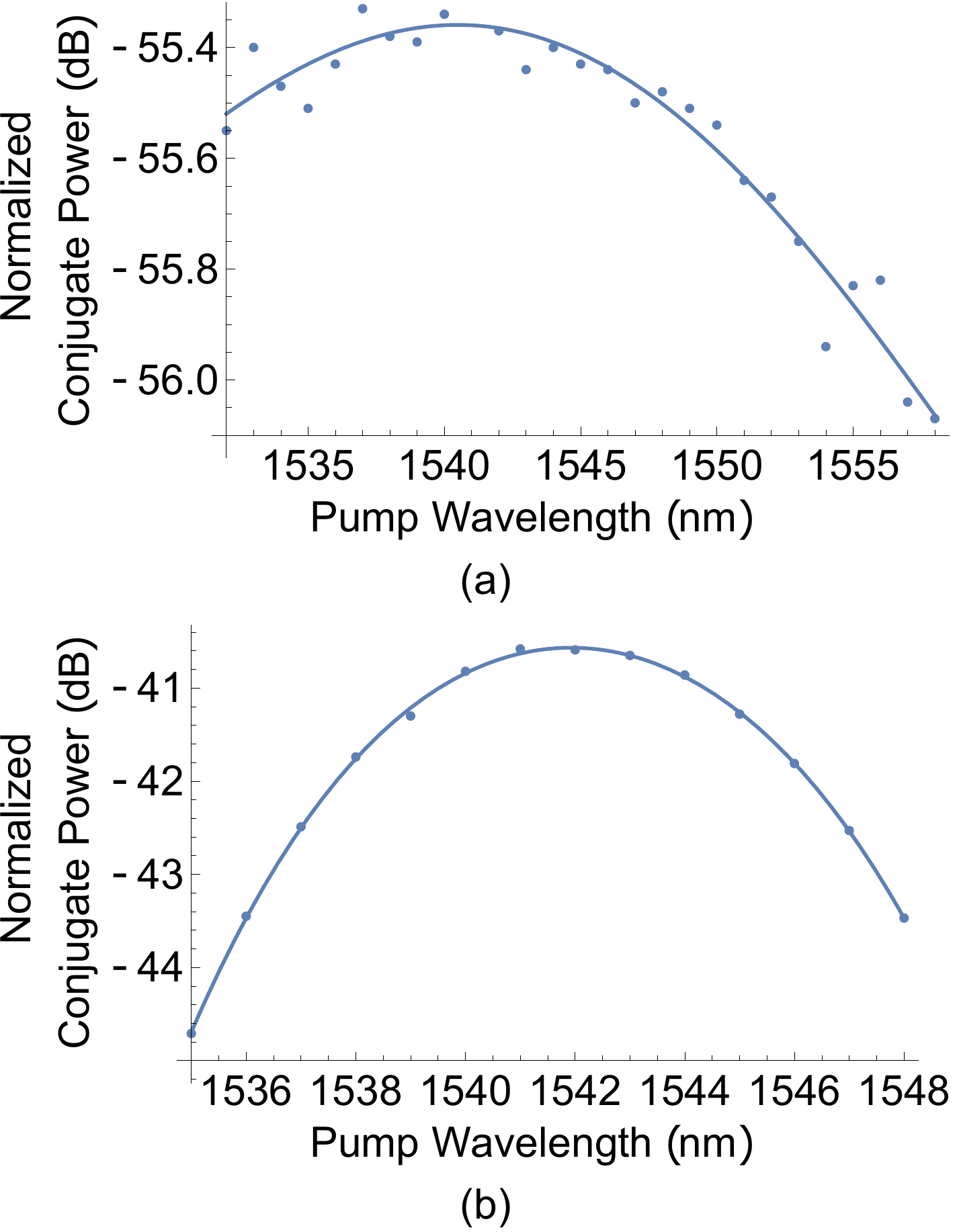}}
\caption{Highly nonlinear fiber ZDW measurement. (a) 98-m HNLF. Peak extracted from curve fit is $1540.48\pm0.06$ nm. (b) 510-m HNLF. Peak extracted from curve fit is $1541.93\pm0.04$ nm.}
\label{fig:HNLFZDW}
\end{figure}

The normalized conjugate power from FWM is shown in Fig.~\ref{fig:HNLFZDW}a for the 98-m HNLF and Fig.~\ref{fig:HNLFZDW}b for the 510-m HNLF. We then use least-squares fitting algorithm~\cite{Mathematica} to fit the data to a quartic polynomial and extract the peak wavelength (ZDW) with error bars from the fit. The extracted ZDW is $1540.48\pm0.06$ nm and $1541.93\pm0.04$ for the 98-m and 510-m HNLF spools, respectively. Using the curve fit allows for a precise determination of the average ZDW. This measurement is a necessary first step so that further measurements can have the pump centered at the average ZDW for optimal FWM.

\subsection{Zero-dispersion-wavelength variation}
Four-wave mixing is most efficient when the pump is at the ZDW; if the ZDW were to vary along the length of the fiber significantly, that would decrease the FWM efficiency~\cite{Karlsson:98}. Although HNLF was made to address the issue of varying ZDW by having a higher non-linearity so less fiber can be used~\cite{Andrekson:20}, it still can have ZDW variation itself~\cite{ChavezBoggio:07}, which can be problematic. In fact, a randomly varying ZDW has been shown to degrade the noise figure of a phase-sensitive amplifier by adding noise photons~\cite{Karlsson:98}. Since a FWM-based squeezed-light source can also be used as a phase-sensitive amplifier that means that a randomly varying ZDW will also degrade the squeezing.

\begin{figure}
\centerline{\includegraphics[width=1\columnwidth]{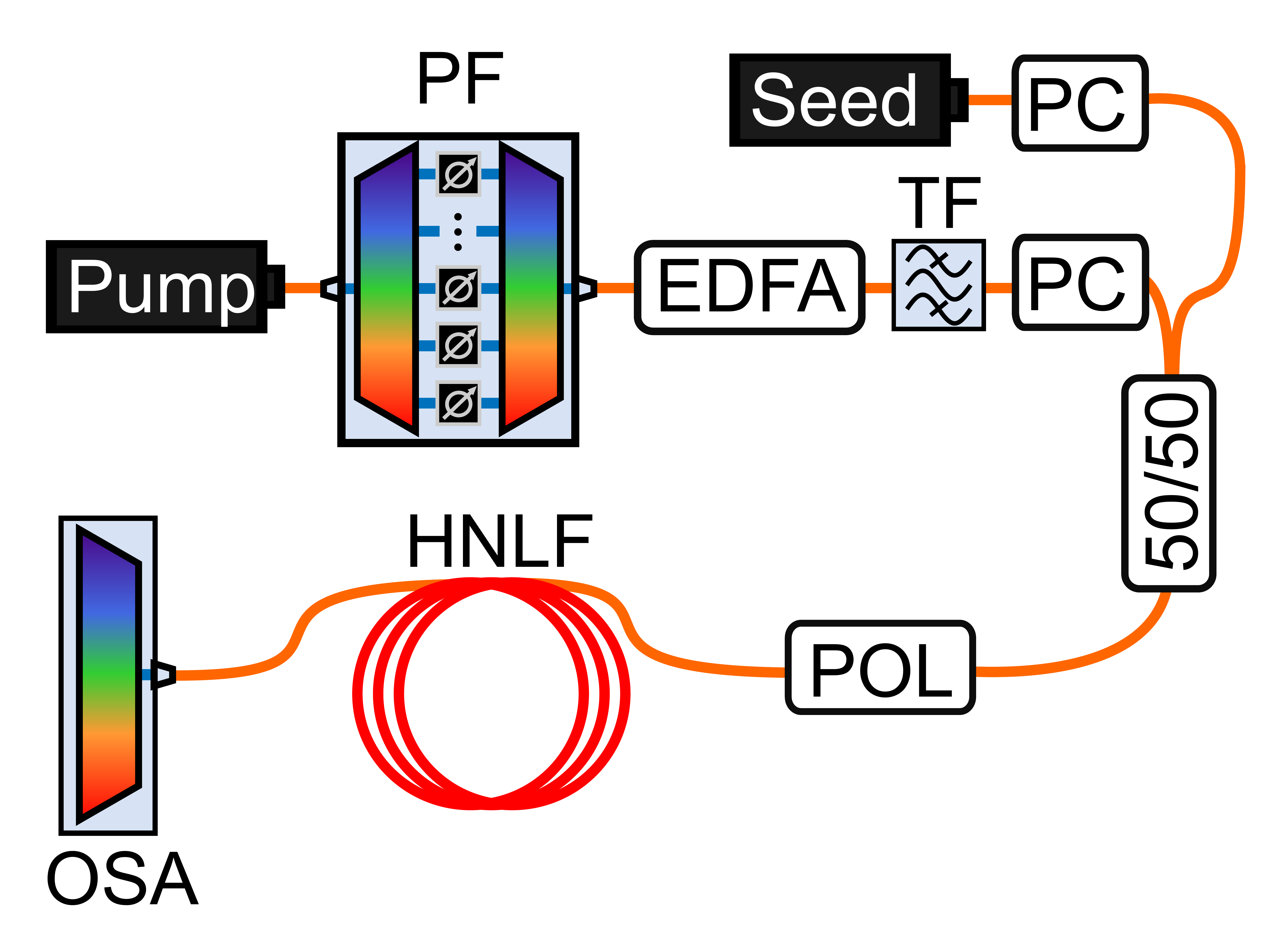}}
\caption{Zero-dispersion-wavelength variation measurement setup. We prepare the pump laser (set at the average ZDW for each measurement) with filtering and amplification and combine it with the seed laser (set at 1471 nm) before co-polarizing them and transmitting them through HNLF. Afterwards the spectra is recorded on the OSA. EDFA = erbium-doped fiber amplifier. HNLF = highly nonlinear fiber. OSA = optical spectrum analyzer. PC = polarization controller. PF = programmable filter. TF = tunable filter.}
\label{fig:HNLFZDWvarsetup}
\end{figure}

To easily characterize ZDW variation in the HNLF, we use a modified version of the method described in Ref.~\cite{ChavezBoggio:07}, shown in Fig.~\ref{fig:HNLFZDWvarsetup}. The method depends on a narrowband seed laser well outside of the phasematching bandwidth (about 50 nm in HNLF~\cite{Andrekson:20}) so that the phasematching condition is very narrow. We use a 5-ps 33-MHz mode-locked fiber laser (Laser-Femto Mercury 1550-005-5000-PM) as the pump laser, centered on the zero-dispersion wavelength. Using a pulsed laser increases the peak power for more efficient four-wave mixing and also has a broad bandwidth to help see if there is ZDW variation within the bandwidth of the fiber-laser pulses (on the order of 1~nm). The pump-laser pulse propagates through a 1-nm bandpass filter (Finisar Waveshaper 1000A) to clean up the laser emission spectrum, then is amplified using an erbium-doped fiber amplifier (Pritel SCG-40) before filtering with a 100-GHz tunable bandpass filter (DiCon Fiberoptics TF-1550-0.8-9/9LT-FC/A-1) to filter out the amplified spontaneous emission. We combine the pump with a narrowband (100-kHz linewidth, broadened to 50-500 MHz) tunable seed laser (Hewlett Packard 8168C) set at 1471~nm. We record the spectra after the HNLF with the OSA (Yokogawa AQ6370B). We use polarization controllers to optimize the transmission through the polarizer before the HNLF.

\begin{figure*}
\centerline{\includegraphics[width=0.75\textwidth]{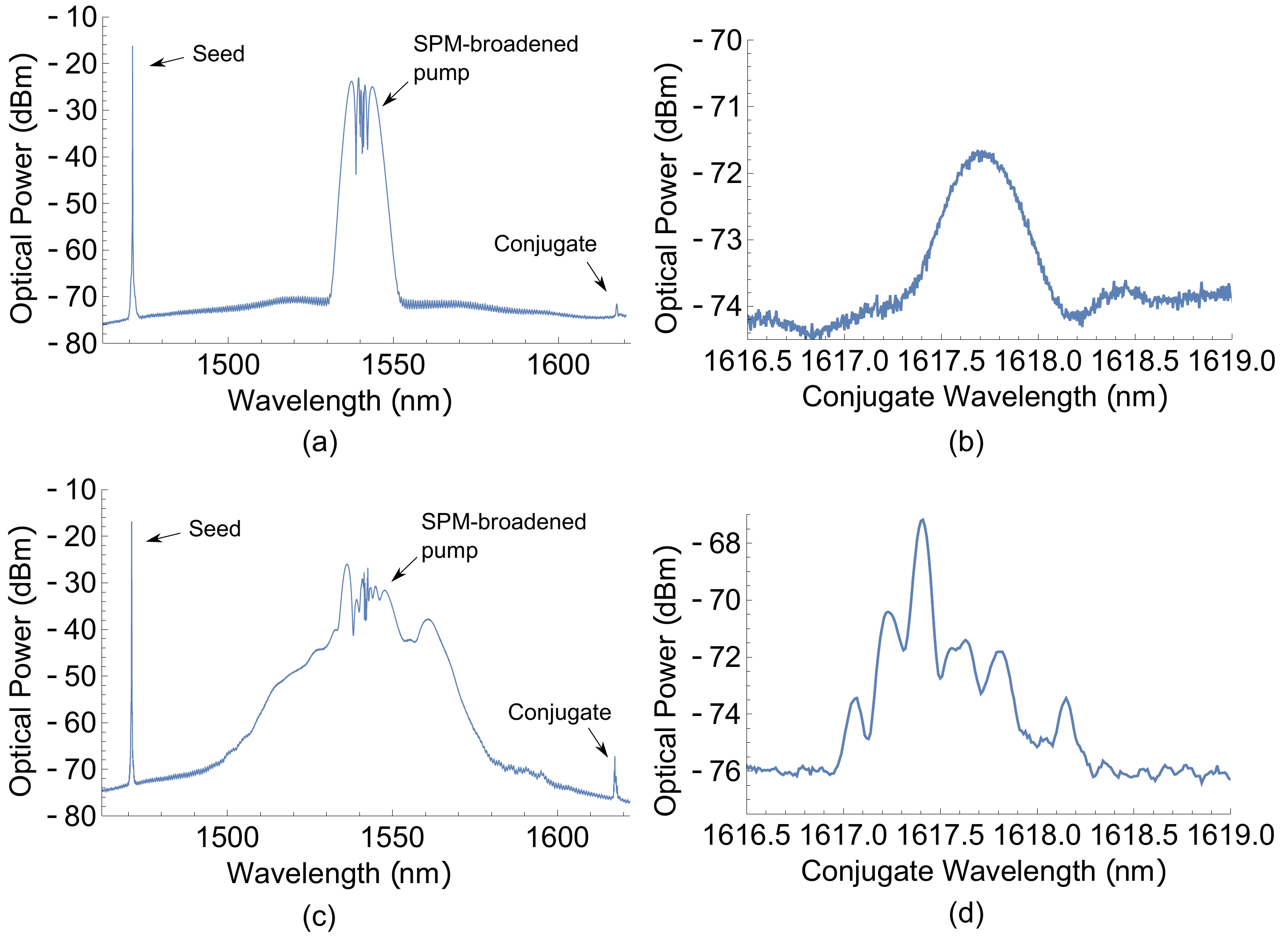}}
\caption{Highly nonlinear fiber ZDW variation measurement. HNLF-output spectra recorded on the OSA.  For all panels the OSA noise floor is about -80 dBm.  (a) 98-m HNLF. The pump was centered at about 1540.5 nm. (b) close-up on conjugate in (a). Resolution bandwidth (RBW) is 0.02 nm  for (a) and (b). (c) 510-m HNLF.  The pump was centered at about 1542 nm. (d) close-up on conjugate in (c). RBW is 0.05 nm for (c) and (d)}.
\label{fig:HNLFZDWvar}
\end{figure*}

Fig.~\ref{fig:HNLFZDWvar} shows the spectra recorded after each HNLF. Notably, in panels (a) and (c), the pump spectra,  initially about 1-nm wide, has been significantly broadened in the HNLF due to self-phase modulation, which could also be interpreted as degenerate four-wave mixing when the pump is at the ZDW~\cite{Marhic_1991}. Nevertheless, the conjugate four-wave mixing product between the pump and seed can be seen in each panel. Fig.~\ref{fig:HNLFZDWvar}(b) and Fig.~\ref{fig:HNLFZDWvar}(d) provide a close up of the conjugate spectra emitted from each fiber. There is also a slight difference in the background level between the 98-m and 510-m HNLF. The difference in length, and therefore attenuation and non-linear strength, is the most likely reason for this difference since all other aspects of the experiment were held constant.

In Fig.~\ref{fig:HNLFZDWvar}(b), there does not appear to be appreciable ZDW variation in the 98-m HNLF since there is only one dominant peak whereas there are several distinct peaks when using the 510-m HNLF in Fig.~\ref{fig:HNLFZDWvar}(d). The peaks are expected to be narrower for the 510-m fiber since the peak width scales as $1/L$, where $L$ is the fiber length~\cite{ChavezBoggio:07}. ZDW variation primarily adds noise to wavelengths far away from the ZDW; the added noise near the zero-dispersion wavelength, from ZDW variation similar to that of the 510-m HNLF, appears to be minimal~\cite{1632256}. With these facts in mind, we do not present a model for zero-dispersion wavelength variation's affect on squeezing; doing so would be non-trivial due to the stochastic differential equations involved. For the interested reader, this could likely be done by adapting the analysis of Ref.~\cite{1632256} to make a simplectic map for two-mode non-degenerate four-wave mixing.

\subsection{Polarization-noise characterization}
Most analyses of squeezing assume a single polarization and that the LO and signal have the same polarization in homodyne detection. Thus, when projecting onto a single polarization, e.g., when there is an inline polarizer or interference at a beamsplitter, polarization noise will transform to intensity noise. {In the case of interference at a beamsplitter, we would expect reduced interference visibility at the beamsplitter from polarization noise because of the noisily reduced polarization overlap of the interfering beams and a noisier interference output because this overlap is affected by a noise process.} On the other hand, reduced overlap of the LO and signal in homodyne detection due to polarization noise, will be an effective loss on the detected signal. A common source of polarization noise is Raman scattering. The non-zero response time of the $\chi^{(3)}$ nonlinearity leads to coupling to the vibration phonon bath, causing the addition of excess Raman noise. But near the ZDW wavelength the added noise is small for modest non-linear phase shifts~\cite{1632256}. On the other hand, a relatively unnoticed nonlinear effect in fiber that could be a major source of noise in non-polarization maintaining fiber is the nonlinear polarization mode dispersion~\cite{Wai:96,Wai:97}. At low intensities this can be simply described by different wavelengths acquiring slightly different nonlinear phase shifts~\cite{Wai:95}. At higher intensities, the dynamics are not as simple and have been shown to add excess noise~\cite{Wai:95,Wai:96,Wai:97}. If this noise source is large enough, this has potential to degrade squeezed light generation because, for example, in the symmetric Sagnac proposal, the pump is split into two paths, clockwise and counter-clockwise propagation. Non-linear polarization noise will not be instantaneously identical in these paths, though the average may be identical, thus we would expect to have noisier and imperfect interference at the beamsplitter. From this, we would expect polarization noise to produce excess noise that would compete with any possible squeezed vacuum in the dark port.
\begin{figure}
\centerline{\includegraphics[width=1\columnwidth]{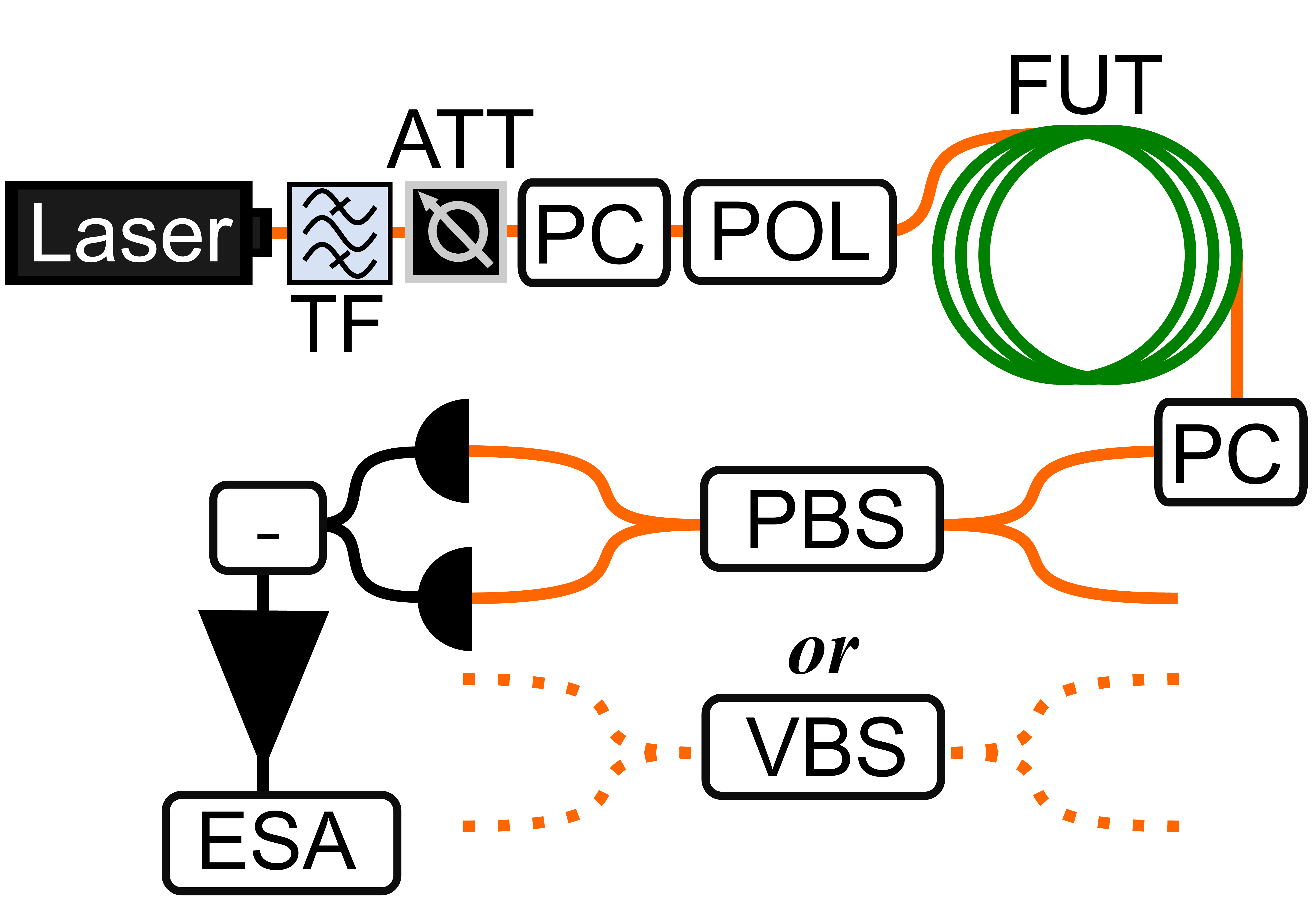}}
\caption{Polarization noise measurement. A pulsed laser is prepared via filtering, attenuation, and polarization then launched into the fiber under test. The output is coupled into a balanced detector with either a fiber-coupled polarizing beamsplitter or a variable fiber beamsplitter. The amplified detector is read out by an electronic spectrum analyzer. ATT = Adjustable attenuator. ESA = electronic spectrum analyzer. PBS = polarizing beamsplitter. PC = polarization controller. POL = polarizer. TF = tunable filter. VBS = variable beamsplitter.}
\label{fig:polnoisesetup}
\end{figure}

To measure any added polarization noise, shown in Fig.~\ref{fig:polnoisesetup} we use a $\tau_{{\text{FWHM}}}=5$ ps {FWHM}, $R=33$ MHz mode-locked fiber laser (Laser-Femto) and prepare it by going through several bandpass filters (Pritel TFA-20 and TFA-40 and DiCon Fiberoptics TF-1550-0.8-9/9LT-FC/A-1), an adjustable attenuator, and an inline fiber polarizer. After preparation, the laser propagates through the fiber under test (FUT) and is either split by a variable fiber beamsplitter (VBS, Newport F-CPL-1550-N-FA) or a fiber-coupled polarizing beamsplitter (PBS, Oz Optics FOBS-22P-1111-9/125-SSSS-1550-PBS-60-3A3A3A3A-1-1-H) and the outputs are sent to an amplified balanced InGaAs photodiode pair (Thorlabs PDB430C, modified as in Ref.~\cite{Chapman:22}) which is connected to a electronic spectrum analyzer (ESA, Agilent N9000A CXA Signal Analyzer). We measure the noise in zero-span mode at 6~MHz using a resolution bandwidth of 510 kHz and a video bandwidth of 30 Hz. 6 MHz was chosen because it is within our detector bandwidth and is not at a guided acoustic-wave Brillouin scattering frequency. The VBS is tuned for maximal common-mode rejection ratio and, when in use, enables measurement of the shot noise for comparison to the polarization noise. To make sure the noise measurements are comparable, we measure the input power going into one of the photodiodes (about 50 $\mu$W) with the VBS and PBS after optimizing the common-mode rejection ratio {to about 40-50 dB} (by either tuning the splitting ratio for the VBS or tuning the polarization for the PBS) and change the adjustable attenuator so they are the same. We have verified the linearity of the detector by comparing the measured shot-noise power for a range of input powers and are working in that linear regime. Furthermore, we have verified the accuracy of the VBS measurement compared to the theoretical shot noise and found them to agree within about 0.2 dB; see Appendix A for more information.

\begin{table*}
\caption{Summary of polarization noise measurements. The fourth column is calculated by subtracting the second from the third column.  The error on the noise measurements is about 0.5 dB. The electronics noise is at -88.2 dBm. FUT (Fiber under test). SMF (single-mode fiber). PMF (polarization-maintaining fiber). HNLF (highly nonlinear fiber).}
\label{tab:polnoise}
\centering
\begin{tabular}{c c c c c}
\hline
FUT &Shot noise (dBm) & Noise using PBS (dBm) & Polarization noise (dB)\\
\hline
None & -82.5 & -81.8 & 0.7\\
100-m SMF & -83 &-79.3 &3.7\\
98-m HNLF &-84.1&-76.4 &7.7\\
510-m HNLF &84.7 &-71.6 & 13.1\\
1-m PMF & -83.2 &-82.4 &0.8\\
22-m PMF&-84.05 &-83.9 &0.15\\
78-m PMF &-84.05 &-83.85&0.2\\
\hline
\end{tabular}
\end{table*}

In Table~\ref{tab:polnoise}, for each FUT we show the measured shot noise and noise measured using the PBS. We also calculate the polarization noise by subtracting the shot noise from the noise measured using the PBS. In our study, we did not control the input polarization to each FUT, except the input polarization into PMF was along the slow axis, as is typical. Here we see that there is significant polarization noise for all long lengths of non-polarization maintaining fiber and the noise is even greater for the highly nonlinear fiber, as expected due to its higher nonlinearity. As Table~\ref{tab:polnoise} shows, the non-polarizing beamsplitter (the VBS) before the balanced detector is not sensitive to this polarization noise so the LO in homodyne detection should register the shot noise, assuming the LO has sufficiently low technical noise and the homodyne detection has sufficiently high common-mode rejection ratio, as we have confirmed. 
As described at the beginning of this section, we expect polarization noise to manifest primarily as intensity noise when there involves projection onto a single polarization or interference. Below we model this polarization noise in an example squeezing configuration to simulate its effect.

Here we model the symmetric Sagnac and added noise using Gaussian matrices introduced above. The fiber provides a distributed (along the fiber) source of squeezing and noise. Thus to say that the squeezing is generated and then the noise is added or vice versa could lead to erroneous conclusions. To counteract this, but work within the constraints of the simplectic matrices, we have chosen to discretize the propagation into 10 segments where each provides some squeezing then adds noise. Since the interferometer has two paths, we have $N=2$ bosonic modes. Thus, we will use the 2-mode beamsplitter introduced in \eqref{eq:BS}. Moreover, since each path has the non-linear fiber, we introduce the 2 (independent) mode FWM and Kerr operators simplectic matrices:
\begin{equation}
    \textbf{S}_{2\text{FWM}}=\begin{pmatrix}
    \textbf{S}_{\text{FWM}} & \textbf{0}\\
    \textbf{0} & \textbf{S}_{\text{FWM}}\\
    \end{pmatrix}
\end{equation}
and
\begin{equation}
    \textbf{S}_{2K}=\begin{pmatrix}
    \textbf{S}_{K} & \textbf{0}\\
    \textbf{0} & \textbf{S}_{K}\\
    \end{pmatrix},
\end{equation}
where \textbf{0} is the 0 matrix. We also model detection loss at the end using the Gaussian channel with $\textbf{T}_l=\sqrt{\eta}\textbf{I}$ and $\textbf{N}_l=(1-\eta)\textbf{I}$, in this case using $4\times4$ matrices to accommodate the two modes.

\begin{figure*}
\centerline{\includegraphics[width=0.75\textwidth]{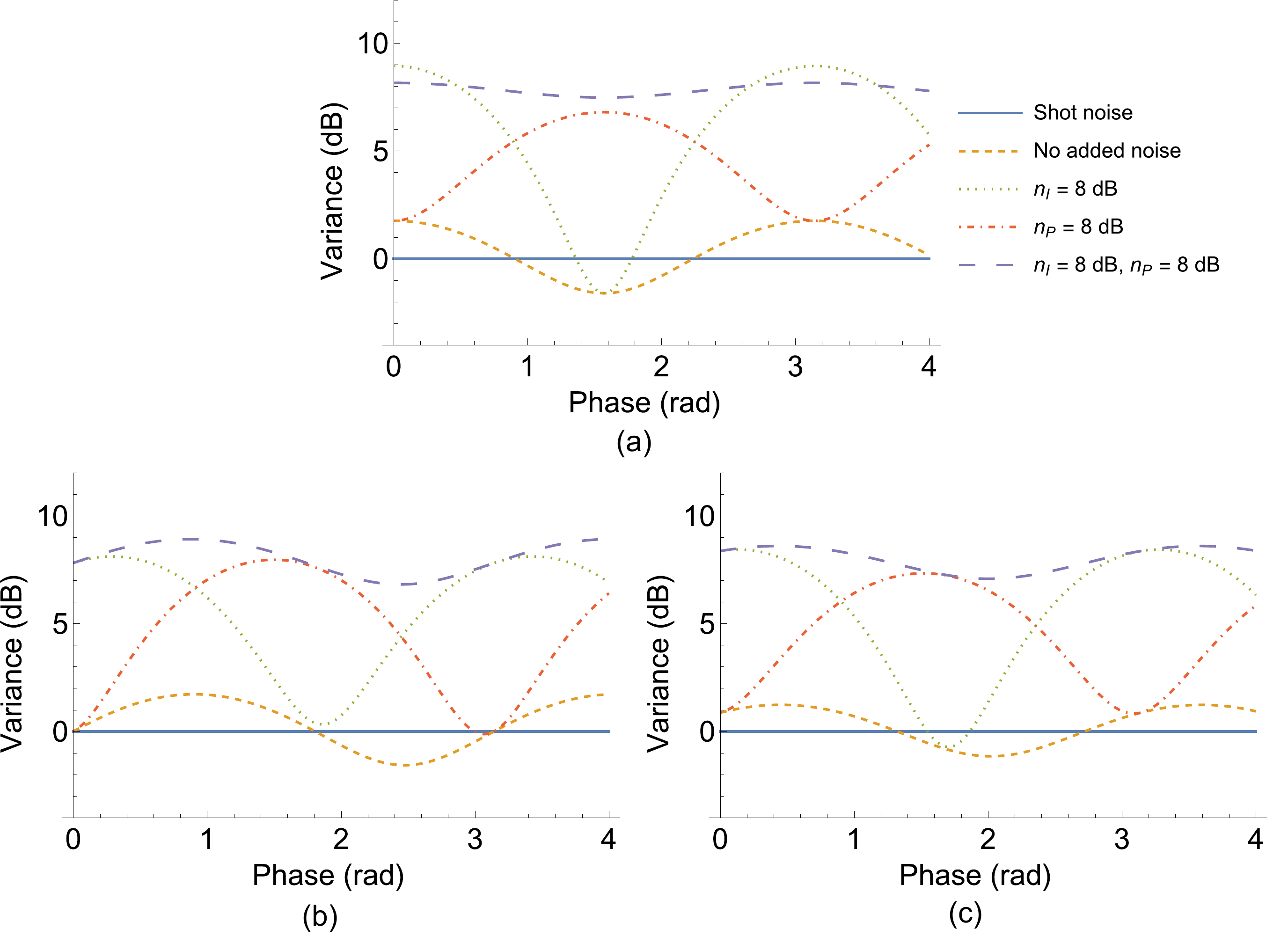}}
\caption{Symmetric Sagnac squeezing with simulated noise. (a) 10 segments of FWM squeezing using {$G=1.00059$}. (b) 10 segments of Kerr squeezing using {$r_K=0.024$}. (c) 10 segments of FWM squeezing and Kerr squeezing using {$G=1.00015$ and $r_K=0.012$}. All panels used $\eta=0.8$ and share legend.}
\label{fig:polnoisesim}
\end{figure*}

For example, the model for 2 segments looks like:

 \begin{align}
     &\textbf{V}_{\text{Sagnac}}=\textbf{T}_l\textbf{B}(1/2)\notag\\
     &\times\bigg( \textbf{S}_{\text{2K}}\textbf{S}_{\text{2FWM}}\big( \textbf{S}_{\text{2K}}\textbf{S}_{\text{2FWM}}\textbf{B}(1/2)\textbf{V}_\textbf{2p}\textbf{B}(1/2)^T\notag\\
     &\times\textbf{S}_{\text{2FWM}}^T\textbf{S}_{\text{2K}}^T+\textbf{N}_{{2}}\big)\textbf{S}_{\text{2FWM}}^T\textbf{S}_{\text{2K}}^T+\textbf{N}_{{2}}\bigg)\textbf{B}(1/2)^T\textbf{T}^T_l+\textbf{N}_l,
 \end{align}
where $\textbf{V}_\textbf{2p}$ is the two-mode version of $\textbf{V}_\textbf{p}$ [defined in \eqref{eq:pump}] and $\textbf{N}_2$ is the two-mode version of $\textbf{N}$ [defined in {\eqref{eq:N}}]. To simulate homodyne detection, we look at a single quadrature and use the rotation matrix to simulate the LO phase. Due to the phase convention, Output mode 2 is the dark port where we expect to see squeezing. In Fig.~\ref{fig:polnoisesim}, we plot the intensity quadrature of output mode 2 (element [3,3]) of $\textbf{V}_{\text{Sagnac}}$, rotated by LO phase for several different squeezing and noise scenarios. 

We model the polarization noise as intensity noise but we also show the effect of added phase noise of the same magnitude. From Table~\ref{tab:polnoise}, the polarization noise of the 98-m HNLF is about 8 dB. We model this by breaking it into 10 segments along with the squeezing, which we model by also being pumped by the same laser at the same average power ($\Bar{P}=50$ $\mu$W). We model three squeezing scenarios: FWM, Kerr, FWM and Kerr (as if half the segment length went to each). We model the final scenario because, when centered at the ZDW, a high-energy broadband pulse is likely to have enough power and bandwidth to exhibit both effects simultaneously. {10 segments were chosen to balance between computation time and to have enough segments that the result is not dominated by the order of the three processes (Kerr, FWM, and added noise). For small n, close to 1, the results varied significantly depending on the order of the three processes.} Both FWM and Kerr squeezing are applicable for our type-II HNLF with the ZDW in the 1550-nm region. In fact, FWM is expected for any type of HNLF whenever the pump laser is near the ZDW or in the anomalous dispersion regime which allows for phase matching, otherwise Kerr squeezing will dominate. Using the pulse peak power, we calculate {$r_K=\gamma P_p L=0.024$ and $G=1+(\gamma P_p L)^2=1.00059$ for $\gamma=9.8$ (W km)$^{-1}$, $P_p=0.88 \Bar{P}/(R\tau_{{\text{FWHM}}})=0.266$ {W}, and
$L=98/10=9.8$ m. For the half and half scenario, we calculate $r_K=0.012$ and $G=1.00015$.} These values may look small (no squeezing is $r_K=0$ and $G=1$), but as seen in Fig.~\ref{fig:polnoisesim}, they can generate an appreciable amount of squeezing when there's 10 segments even with a low, but still realistic, amount of loss ($\eta=0.8$).

FWM squeezing is phase squeezing so the polarization noise, modeled as intensity noise in Fig.~\ref{fig:polnoisesim}(a) does not degrade the squeezing but adds to the anti-squeezing. FWM is very sensitive to phase noise and that is seen in Fig.~\ref{fig:polnoisesim}(a) by the absence of squeezing with the same amount of excess phase noise. Kerr squeezing on the other hand is squeezed in between the quadratures, tending toward intensity squeezing as $r_K$ increases. As such, in Fig.~\ref{fig:polnoisesim}(b), we see that Kerr squeezing is affected by the intensity (polarization) noise and by phase noise, but more so by the intensity noise. When Kerr and FWM squeezing are combined [Fig.~\ref{fig:polnoisesim}(c)], the squeezing dip is at a phase in between their dips by themselves. This combination is sensitive to both types of noise, but the FWM appears to dominate in this 50/50 combination because it is more sensitive to phase noise. Overall, we find that polarization noise, insofar as it can be modeled by intensity noise, can degrade Kerr squeezing and combinations of FWM and Kerr squeezing.

\section{Conclusion}
We have characterized two lengths of non-polarization-maintaining HNLF and found several sources of excess noise. Nonlinear PMD in particular is a significant source of excess noise. We model the polarization noise in an example system for several squeezing configurations and show that Kerr squeezing is susceptible to polarization noise while FWM is not, except for more noise added to the anti-squeezing.
While it may be worth considering a polarization-maintaining version of this fiber, which may still be improved for Kerr-squeezed-light generation, it comes with a significantly increased cost compared to normal polarization-maintaining fiber  and even compared to non-polarization-maintaining HNLF. In situations with very limited pump power or limited choices of pump wavelength, the increased non-linearity may be worth the extra cost.

\section*{Appendix A: detector calibration and theoretical shot-noise comparison}\label{app:SNRIN}
\subsection{Detector Calibration}
To calibrate the balanced detector, we measure the responsivity of each photodiode by desoldering the lead connected to the amplifier and measuring the DC current flowing through the diode using a multi-meter and a calibrated optical power meter (Thorlabs PM100D).  We divide the current and optical power to get the responsivity. The responsivity can be expressed as 
\begin{equation}
\mathcal{R}=\eta \frac{e}{h f},
\end{equation}
where $\eta$ is the quantum efficiency, $e$ is the elementary charge, $h$ is Planck's constant, and $f$ is the optical frequency. The responsivity has units of A/W. This formula allows us to know the detector's quantum efficiency which is important for squeezed light measurements as such measurements are very sensitive to loss. At 1550 nm, most detectors are made of InGaAs and have a responsivity of about 1 A/W, which is what we measured for the detectors we used.

We also calibrated the gain $\mathcal{R}G$ (V/W) of each balanced detector by using a calibrated power meter (Thorlabs PM100D) to measure the input power and a multi-meter to measure the DC voltage across a 50-Ohm load connected to the output of the amplified detector. In the modified PDB430C, we measured 118.4 mV for a 35.7 $\mu$W input giving
\begin{equation}
 \mathcal{R}G=\frac{0.1184}{0.0000357}=3316.
 \end{equation}
\subsection{Shot-noise and relative-intensity noise}
To compare the theoretical shot-noise level with the measured shot noise, we found it easiest to calculate the relative-intensity noise of both and compare them.

The optical relative intensity noise due to shot noise is simply~\cite{paschotta2004encyclopedia}
\begin{equation}
RIN_{\text{SN}}(\omega)=\frac{2hf}{ P_{\text{opt,avg}}},
\end{equation}
where $P_{\text{opt,avg}}$ is the average optical power. Due to the square-law detection principle of photodiodes, the general detected optical relative intensity noise can be expressed simply as (from Eq. (3.1.4) in Ref.~\cite{hui2009fiber})
\begin{equation}
RIN(\omega)=\frac{S_p(\omega)}{(\mathcal{R} P_{\text{opt,avg}})^2},
\label{RIN1}
\end{equation}
where $S_p(\omega)$ is the measured electrical noise power spectral density.
In the case of amplified detectors, \eqref{RIN1} is transformed into
\begin{equation}
RIN(\omega)=\frac{S_p(\omega) R_{l}}{(\mathcal{R} G P_{\text{opt,avg}})^2},
\label{RIN2}
\end{equation}
where $R_{l}$ is the load resistance (nominally 50 Ohms) and G is the amplifier gain (V/A). As discussed above, it is often easier to measure $\mathcal{R} G$ (V/W) directly and that is what we have done. Using $10\text{Log}_{10}(x)$ to convert \eqref{RIN2} to the dB scale would result in the common units of dBc/Hz.

\begin{backmatter}

\bmsection{Acknowledgments}
The authors acknowledge Raphael Pooser for useful discussions, including his advice on shot-noise calibration and electronic spectrum analyzer configuration.
This work was performed at Oak Ridge National Laboratory, operated by UT-Battelle for the U.S. Department of Energy under contract no. DE-AC05-00OR22725.
Funding was provided by the U.S. Department of Energy, Office of Science, Advanced Scientific Computing Research Program Office, through the Transparent Optical Quantum Networks for Distributed Science Program (Field Work Proposal ERKJ355).\\
\bmsection{Disclosures}
The authors declare no conflicts of interest.

\bmsection{Data Availability Statement}
Data underlying the results presented in this paper are not publicly available at this time but may be obtained from the authors upon reasonable request.

\end{backmatter}


\begin{thebibliography}{10}
\newcommand{\enquote}[1]{``#1''}

\bibitem{bachor2019guide}
H.-A. Bachor and T.~C. Ralph, \emph{A guide to experiments in quantum optics}
  (John Wiley \& Sons, 2019).

\bibitem{QSRev}
B.~J. Lawrie, P.~D. Lett, A.~M. Marino, and R.~C. Pooser, \enquote{Quantum
  sensing with squeezed light,} {\protect\JournalTitle{ACS Photonics}}
  \textbf{6}, 1307--1318 (2019).

\bibitem{Aasi2013}
J.~Aasi, J.~Abadie, B.~P. Abbott, R.~Abbott, T.~D. Abbott, M.~R. Abernathy,
  C.~Adams, T.~Adams, P.~Addesso, R.~X. Adhikari, C.~Affeldt, O.~D. Aguiar,
  P.~Ajith, B.~Allen, E.~Amador~Ceron, D.~Amariutei, S.~B. Anderson, W.~G.
  Anderson, K.~Arai, M.~C. Araya, C.~Arceneaux, S.~Ast, S.~M. Aston,
  D.~Atkinson, P.~Aufmuth, C.~Aulbert, L.~Austin, B.~E. Aylott, S.~Babak, P.~T.
  Baker, S.~Ballmer, Y.~Bao, J.~C. Barayoga, D.~Barker, B.~Barr, L.~Barsotti,
  M.~A. Barton, I.~Bartos, R.~Bassiri, J.~Batch, J.~Bauchrowitz, B.~Behnke,
  A.~S. Bell, C.~Bell, G.~Bergmann, J.~M. Berliner, A.~Bertolini,
  J.~Betzwieser, N.~Beveridge, P.~T. Beyersdorf, T.~Bhadbhade, I.~A. Bilenko,
  G.~Billingsley, J.~Birch, S.~Biscans, E.~Black, J.~K. Blackburn,
  L.~Blackburn, D.~Blair, B.~Bland, O.~Bock, T.~P. Bodiya, C.~Bogan, C.~Bond,
  R.~Bork, M.~Born, S.~Bose, J.~Bowers, P.~R. Brady, V.~B. Braginsky, J.~E.
  Brau, J.~Breyer, D.~O. Bridges, M.~Brinkmann, M.~Britzger, A.~F. Brooks,
  D.~A. Brown, D.~D. Brown, K.~Buckland, F.~Br{\"u}ckner, B.~C. Buchler,
  A.~Buonanno, J.~Burguet-Castell, R.~L. Byer, L.~Cadonati, J.~B. Camp,
  P.~Campsie, K.~Cannon, J.~Cao, C.~D. Capano, L.~Carbone, S.~Caride, A.~D.
  Castiglia, S.~Caudill, M.~Cavagli{\`a}, C.~Cepeda, T.~Chalermsongsak,
  S.~Chao, P.~Charlton, X.~Chen, Y.~Chen, H.-S. Cho, J.~H. Chow,
  N.~Christensen, Q.~Chu, S.~S.~Y. Chua, C.~T.~Y. Chung, G.~Ciani, F.~Clara,
  D.~E. Clark, J.~A. Clark, M.~Constancio~Junior, D.~Cook, T.~R. Corbitt,
  M.~Cordier, N.~Cornish, A.~Corsi, C.~A. Costa, M.~W. Coughlin, S.~Countryman,
  P.~Couvares, D.~M. Coward, M.~Cowart, D.~C. Coyne, K.~Craig, J.~D.~E.
  Creighton, T.~D. Creighton, A.~Cumming, L.~Cunningham, K.~Dahl, M.~Damjanic,
  S.~L. Danilishin, K.~Danzmann, B.~Daudert, H.~Daveloza, G.~S. Davies, E.~J.
  Daw, T.~Dayanga, E.~Deleeuw, T.~Denker, T.~Dent, V.~Dergachev, R.~DeRosa,
  R.~DeSalvo, S.~Dhurandhar, I.~Di~Palma, M.~D{\'i}az, A.~Dietz, F.~Donovan,
  K.~L. Dooley, S.~Doravari, S.~Drasco, R.~W.~P. Drever, J.~C. Driggers, Z.~Du,
  J.-C. Dumas, S.~Dwyer, T.~Eberle, M.~Edwards, A.~Effler, P.~Ehrens, S.~S.
  Eikenberry, R.~Engel, R.~Essick, T.~Etzel, K.~Evans, M.~Evans, T.~Evans,
  M.~Factourovich, S.~Fairhurst, Q.~Fang, B.~F. Farr, W.~Farr, M.~Favata,
  D.~Fazi, H.~Fehrmann, D.~Feldbaum, L.~S. Finn, R.~P. Fisher, S.~Foley,
  E.~Forsi, N.~Fotopoulos, M.~Frede, M.~A. Frei, Z.~Frei, A.~Freise, R.~Frey,
  T.~T. Fricke, D.~Friedrich, P.~Fritschel, V.~V. Frolov, M.-K. Fujimoto, P.~J.
  Fulda, M.~Fyffe, J.~Gair, J.~Garcia, N.~Gehrels, G.~Gelencser, L.~{\'A}.
  Gergely, S.~Ghosh, J.~A. Giaime, S.~Giampanis, K.~D. Giardina,
  S.~Gil-Casanova, C.~Gill, J.~Gleason, E.~Goetz, G.~Gonz{\'a}lez, N.~Gordon,
  M.~L. Gorodetsky, S.~Gossan, S.~Go{\ss}ler, C.~Graef, P.~B. Graff, A.~Grant,
  S.~Gras, C.~Gray, R.~J.~S. Greenhalgh, A.~M. Gretarsson, C.~Griffo, H.~Grote,
  K.~Grover, S.~Grunewald, C.~Guido, E.~K. Gustafson, R.~Gustafson, D.~Hammer,
  G.~Hammond, J.~Hanks, C.~Hanna, J.~Hanson, K.~Haris, J.~Harms, G.~M. Harry,
  I.~W. Harry, E.~D. Harstad, M.~T. Hartman, K.~Haughian, K.~Hayama,
  J.~Heefner, M.~C. Heintze, M.~A. Hendry, I.~S. Heng, A.~W. Heptonstall,
  M.~Heurs, M.~Hewitson, S.~Hild, D.~Hoak, K.~A. Hodge, K.~Holt, M.~Holtrop,
  T.~Hong, S.~Hooper, J.~Hough, E.~J. Howell, V.~Huang, E.~A. Huerta,
  B.~Hughey, S.~H. Huttner, M.~Huynh, T.~Huynh-Dinh, D.~R. Ingram, R.~Inta,
  T.~Isogai, A.~Ivanov, B.~R. Iyer, K.~Izumi, M.~Jacobson, E.~James, H.~Jang,
  Y.~J. Jang, E.~Jesse, W.~W. Johnson, D.~Jones, D.~I. Jones, R.~Jones, L.~Ju,
  P.~Kalmus, V.~Kalogera, S.~Kandhasamy, G.~Kang, J.~B. Kanner, R.~Kasturi,
  E.~Katsavounidis, W.~Katzman, H.~Kaufer, K.~Kawabe, S.~Kawamura, F.~Kawazoe,
  D.~Keitel, D.~B. Kelley, W.~Kells, D.~G. Keppel, A.~Khalaidovski, F.~Y.
  Khalili, E.~A. Khazanov, B.~K. Kim, C.~Kim, K.~Kim, and N.~Kim,
  \enquote{Enhanced sensitivity of the {LIGO} gravitational wave detector by
  using squeezed states of light,} {\protect\JournalTitle{Nature Photonics}}
  \textbf{7}, 613--619 (2013).

\bibitem{RevModPhys.77.513}
S.~L. Braunstein and P.~van Loock, \enquote{Quantum information with continuous
  variables,} {\protect\JournalTitle{Rev. Mod. Phys.}} \textbf{77}, 513--577
  (2005).

\bibitem{RevModPhys.84.621}
C.~Weedbrook, S.~Pirandola, R.~Garc\'{\i}a-Patr\'on, N.~J. Cerf, T.~C. Ralph,
  J.~H. Shapiro, and S.~Lloyd, \enquote{Gaussian quantum information,}
  {\protect\JournalTitle{Rev. Mod. Phys.}} \textbf{84}, 621--669 (2012).

\bibitem{PhysRevLett.55.2409}
R.~E. Slusher, L.~W. Hollberg, B.~Yurke, J.~C. Mertz, and J.~F. Valley,
  \enquote{Observation of squeezed states generated by four-wave mixing in an
  optical cavity,} {\protect\JournalTitle{Phys. Rev. Lett.}} \textbf{55},
  2409--2412 (1985).

\bibitem{PhysRevLett.57.691}
R.~M. Shelby, M.~D. Levenson, S.~H. Perlmutter, R.~G. DeVoe, and D.~F. Walls,
  \enquote{Broad-band parametric deamplification of quantum noise in an optical
  fiber,} {\protect\JournalTitle{Phys. Rev. Lett.}} \textbf{57}, 691--694
  (1986).

\bibitem{PhysRevLett.59.2566}
R.~E. Slusher, P.~Grangier, A.~LaPorta, B.~Yurke, and M.~J. Potasek,
  \enquote{Pulsed squeezed light,} {\protect\JournalTitle{Phys. Rev. Lett.}}
  \textbf{59}, 2566--2569 (1987).

\bibitem{Bergman:91}
K.~Bergman and H.~A. Haus, \enquote{Squeezing in fibers with optical pulses,}
  {\protect\JournalTitle{Opt. Lett.}} \textbf{16}, 663--665 (1991).

\bibitem{Bergman:94}
K.~Bergman, H.~A. Haus, E.~P. Ippen, and M.~Shirasaki, \enquote{Squeezing in a
  fiber interferometer with a gigahertz pump,} {\protect\JournalTitle{Opt.
  Lett.}} \textbf{19}, 290--292 (1994).

\bibitem{Sudo1990}
S.~Sudo and H.~Itoh, \enquote{Efficient non-linear optical fibres and their
  applications,} {\protect\JournalTitle{Optical and Quantum Electronics}}
  \textbf{22}, 187--212 (1990).

\bibitem{secondHNLF}
M.~Holmes, D.~Williams, and R.~Manning, \enquote{Highly nonlinear optical fiber
  for all optical processing applications,} {\protect\JournalTitle{IEEE
  Photonics Technology Letters}} \textbf{7}, 1045--1047 (1995).

\bibitem{ONISHI1998204}
M.~Onishi, T.~Okuno, T.~Kashiwada, S.~Ishikawa, N.~Akasaka, and M.~Nishimura,
  \enquote{Highly nonlinear dispersion-shifted fibers and their application to
  broadband wavelength converter,} {\protect\JournalTitle{Optical Fiber
  Technology}} \textbf{4}, 204--214 (1998).

\bibitem{806765}
T.~Okuno, M.~Onishi, T.~Kashiwada, S.~Ishikawa, and M.~Nishimura,
  \enquote{Silica-based functional fibers with enhanced nonlinearity and their
  applications,} {\protect\JournalTitle{IEEE Journal of Selected Topics in
  Quantum Electronics}} \textbf{5}, 1385--1391 (1999).

\bibitem{1561390}
M.-J. Li, S.~Li, and D.~Nolan, \enquote{Nonlinear fibers for signal processing
  using optical {Kerr} effects,} {\protect\JournalTitle{Journal of Lightwave
  Technology}} \textbf{23}, 3606--3614 (2005).

\bibitem{1561391}
M.~Takahashi, R.~Sugizaki, J.~Hiroishi, M.~Tadakuma, Y.~Taniguchi, and T.~Yagi,
  \enquote{Low-loss and low-dispersion-slope highly nonlinear fibers,}
  {\protect\JournalTitle{Journal of Lightwave Technology}} \textbf{23},
  3615--3624 (2005).

\bibitem{HNLFreview}
M.~Hirano, T.~Nakanishi, T.~Okuno, and M.~Onishi, \enquote{Silica-based highly
  nonlinear fibers and their application,} {\protect\JournalTitle{IEEE Journal
  of Selected Topics in Quantum Electronics}} \textbf{15}, 103--113 (2009).

\bibitem{1016354}
J.~Hansryd, P.~Andrekson, M.~Westlund, J.~Li, and P.-O. Hedekvist,
  \enquote{Fiber-based optical parametric amplifiers and their applications,}
  {\protect\JournalTitle{IEEE Journal of Selected Topics in Quantum
  Electronics}} \textbf{8}, 506--520 (2002).

\bibitem{el_20030544}
S.~Radic, C.~J. McKinstrie, R.~M. Jopson, J.~C. Centanni, Q.~Lin, and G.~P.
  Agrawal, \enquote{Record performance of parametric amplifier constructed with
  highly nonlinear fibre,} {\protect\JournalTitle{Electronics Letters}}
  \textbf{39}, 838--839(1) (2003).

\bibitem{632764}
S.~Watanabe, S.~Takeda, G.~Ishikawa, H.~Ooi, J.~Nielsen, and C.~Sonne,
  \enquote{Simultaneous wavelength conversion and optical phase conjugation of
  200 {Gb/s} (5/spl times/40 {Gb/s}) {WDM} signal using a highly nonlinear
  fiber four-wave mixer,} in \emph{Integrated Optics and Optical Fibre
  Communications, 11th International Conference on, and 23rd European
  Conference on Optical Communications (Conf. Publ. No.: 448),}  vol.~5 (1997),
  pp. 1--4 vol.5.

\bibitem{el_20030599}
T.~Okuno, M.~Hirano, T.~Kato, M.~Shigematsu, and M.~Onishi, \enquote{Highly
  nonlinear and perfectly dispersion-flattened fibres for efficient optical
  signal processing applications,} {\protect\JournalTitle{Electronics Letters}}
  \textbf{39}, 972--974(2) (2003).

\bibitem{Tamura:01}
K.~R. Tamura and M.~Nakazawa, \enquote{54-fs, 10-{GHz} soliton generation from
  a polarization-maintaining dispersion-flattened dispersion-decreasing fiber
  pulse compressor,} {\protect\JournalTitle{Opt. Lett.}} \textbf{26}, 762--764
  (2001).

\bibitem{1561376}
T.~Miyamoto, M.~Tanaka, J.~Kobayashi, T.~Tsuzaki, M.~Hirano, T.~Okuno,
  M.~Kakui, and M.~Shigematsu, \enquote{Highly nonlinear fiber-based lumped
  fiber {Raman} amplifier for {CWDM} transmission systems,}
  {\protect\JournalTitle{Journal of Lightwave Technology}} \textbf{23},
  3475--3483 (2005).

\bibitem{1650529}
T.~Inoue, H.~Tobioka, K.~Igarashi, and S.~Namiki, \enquote{Optical pulse
  compression based on stationary rescaled pulse propagation in a comblike
  profiled fiber,} {\protect\JournalTitle{Journal of Lightwave Technology}}
  \textbf{24}, 2510--2522 (2006).

\bibitem{doi:10.1063/1.5048198}
J.~M. Lukens, R.~C. Pooser, and N.~A. Peters, \enquote{A broadband fiber-optic
  nonlinear interferometer,} {\protect\JournalTitle{Applied Physics Letters}}
  \textbf{113}, 091103 (2018).

\bibitem{LevensonNLO}
M.~D. Levenson and S.~S. Kano, \emph{Introduction to Nonlinear Optics, 2e.}
  (Academic Press, 1988).

\bibitem{Andrekson:20}
P.~A. Andrekson and M.~Karlsson, \enquote{Fiber-based phase-sensitive optical
  amplifiers and their applications,} {\protect\JournalTitle{Adv. Opt.
  Photon.}} \textbf{12}, 367--428 (2020).

\bibitem{Anashkina:20}
E.~A. Anashkina, A.~V. Andrianov, J.~F. Corney, and G.~Leuchs,
  \enquote{Chalcogenide fibers for kerr squeezing,} {\protect\JournalTitle{Opt.
  Lett.}} \textbf{45}, 5299--5302 (2020).

\bibitem{PhysRevLett.94.203601}
K.~Hirosawa, H.~Furumochi, A.~Tada, F.~Kannari, M.~Takeoka, and M.~Sasaki,
  \enquote{Photon number squeezing of ultrabroadband laser pulses generated by
  microstructure fibers,} {\protect\JournalTitle{Phys. Rev. Lett.}}
  \textbf{94}, 203601 (2005).

\bibitem{McCormick:07}
C.~F. McCormick, V.~Boyer, E.~Arimondo, and P.~D. Lett, \enquote{Strong
  relative intensity squeezing by four-wave mixing in rubidium vapor,}
  {\protect\JournalTitle{Opt. Lett.}} \textbf{32}, 178--180 (2007).

\bibitem{mechels1997accurate}
S.~Mechels, J.~B. Schlager, and D.~L. Franzen, \enquote{Accurate measurements
  of the zero-dispersion wavelength in optical fibers,}
  {\protect\JournalTitle{Journal of Research of the National Institute of
  Standards and Technology}} \textbf{102}, 333 (1997).

\bibitem{Shirasaki:90}
M.~Shirasaki and H.~A. Haus, \enquote{Squeezing of pulses in a nonlinear
  interferometer,} {\protect\JournalTitle{J. Opt. Soc. Am. B}} \textbf{7},
  30--34 (1990).

\bibitem{PhysRevLett.66.153}
M.~Rosenbluh and R.~M. Shelby, \enquote{Squeezed optical solitons,}
  {\protect\JournalTitle{Phys. Rev. Lett.}} \textbf{66}, 153--156 (1991).

\bibitem{Shirasaki:91}
M.~Shirasaki, \enquote{Squeezing performance of a nonlinear symmetric
  {Mach--Zehnder} interferometer using forward degenerate four-wave mixing,}
  {\protect\JournalTitle{J. Opt. Soc. Am. B}} \textbf{8}, 672--680 (1991).

\bibitem{Marhic_1991}
M.~E. Marhic and C.-H. Hsia, \enquote{Optical amplification and squeezed-light
  generation in fibre interferometers performing degenerate four-wave mixing,}
  {\protect\JournalTitle{Quantum Optics: Journal of the European Optical
  Society Part B}} \textbf{3}, 341--358 (1991).

\bibitem{PhysRevLett.81.2446}
S.~Schmitt, J.~Ficker, M.~Wolff, F.~K\"onig, A.~Sizmann, and G.~Leuchs,
  \enquote{Photon-number squeezed solitons from an asymmetric fiber-optic
  {Sagnac} interferometer,} {\protect\JournalTitle{Phys. Rev. Lett.}}
  \textbf{81}, 2446--2449 (1998).

\bibitem{Levandovsky:99}
D.~Levandovsky, M.~Vasilyev, and P.~Kumar, \enquote{Soliton squeezing in a
  highly transmissive nonlinear optical loop mirror,}
  {\protect\JournalTitle{Opt. Lett.}} \textbf{24}, 89--91 (1999).

\bibitem{SIZMANN1999373}
A.~Sizmann and G.~Leuchs, \enquote{V the optical {Kerr} effect and quantum
  optics in fibers,} in \emph{Progress in Optics,}  vol.~39 E.~Wolf, ed.
  (Elsevier, 1999), pp. 373--469.

\bibitem{PhysRevA.63.032312}
A.~S. Holevo and R.~F. Werner, \enquote{Evaluating capacities of bosonic
  gaussian channels,} {\protect\JournalTitle{Phys. Rev. A}} \textbf{63}, 032312
  (2001).

\bibitem{Wang_2001}
L.~J. Wang, C.~K. Hong, and S.~R. Friberg, \enquote{Generation of correlated
  photons via four-wave mixing in optical fibres,}
  {\protect\JournalTitle{Journal of Optics B: Quantum and Semiclassical
  Optics}} \textbf{3}, 346--352 (2001).

\bibitem{Mathematica}
{Wolfram Research, Inc.}, \enquote{Mathematica, {V}ersion 13.0,} Champaign, IL,
  2022.

\bibitem{Karlsson:98}
M.~Karlsson, \enquote{Four-wave mixing in fibers with randomly varying
  zero-dispersion wavelength,} {\protect\JournalTitle{J. Opt. Soc. Am. B}}
  \textbf{15}, 2269--2275 (1998).

\bibitem{ChavezBoggio:07}
J.~M.~C. Boggio and H.~L. Fragnito, \enquote{Simple four-wave-mixing-based
  method for measuring the ratio between the third- and fourth-order dispersion
  in optical fibers,} {\protect\JournalTitle{J. Opt. Soc. Am. B}} \textbf{24},
  2046--2054 (2007).

\bibitem{1632256}
P.~Velanas, A.~Bogris, and D.~Syvridis, \enquote{Impact of dispersion
  fluctuations on the noise properties of fiber optic parametric amplifiers,}
  {\protect\JournalTitle{Journal of Lightwave Technology}} \textbf{24},
  2171--2178 (2006).

\bibitem{Wai:96}
P.~Wai and C.~Menyak, \enquote{Polarization mode dispersion, decorrelation, and
  diffusion in optical fibers with randomly varying birefringence,}
  {\protect\JournalTitle{Journal of Lightwave Technology}} \textbf{14},
  148--157 (1996).

\bibitem{Wai:97}
P.~K.~A. Wai, W.~L. Kath, C.~R. Menyuk, and J.~W. Zhang, \enquote{Nonlinear
  polarization-mode dispersion in optical fibers with randomly varying
  birefringence,} {\protect\JournalTitle{J. Opt. Soc. Am. B}} \textbf{14},
  2967--2979 (1997).

\bibitem{Wai:95}
P.~K.~A. Wai and C.~R. Menyuk, \enquote{Anisotropic diffusion of the state of
  polarization in optical fibers with randomly varying birefringence,}
  {\protect\JournalTitle{Opt. Lett.}} \textbf{20}, 2493--2495 (1995).

\bibitem{Chapman:22}
J.~C. Chapman and N.~A. Peters, \enquote{Heterodyne spectrometer sensitivity
  limit for quantum networking,} {\protect\JournalTitle{Appl. Opt.}}
  \textbf{61}, 5002--5009 (2022).

\bibitem{paschotta2004encyclopedia}
R.~Paschotta and R.~P. Consulting, \emph{Encyclopedia of Laser Physics and
  Technology} (RP Photonics Consulting, 2004).

\bibitem{hui2009fiber}
R.~Hui and M.~O'Sullivan, \emph{Fiber optic measurement techniques} (Academic
  Press, 2009).

\end{thebibliography}

\end{document}